\documentclass[twocolumn,showpacs,preprintnumbers,amsmath,amssymb]{revtex4}

\usepackage{graphicx}
\usepackage{dcolumn}
\usepackage{bm}

\begin{document}
\title{Numerical analysis of the existence and stability of nonlinear excitations in a parametric model of ferromagnetic chain.}
\author{Alain M. Dikand\'e and J. P. Nguenang}
\affiliation{D\'epartement de Physique, Facult\'e des Sciences, Universit\'e de Douala BP 24157 Douala, Cameroun}  
                 
\date{\today}

\begin{abstract}

 A parametrized spin model was recently introduced and intended for one-dimensional ferromagnets with a deformable 
Zeeman energy. This model is revisited and given more realistic interpretation in terms of a model for ferromagnetic systems with 
nonconvex 
anisotropies. A main virtue of the improved form is its exact reduction to the discrete Remoissenet-Peyrard 
model, i.e. a parametrized version of the Takeno-Homma's discrete sine-Gordon model. The spin-wave phase of the improved 
parametrized spin model is investigated assuming both harmonic and anharmonic excitations. Intrinsic-self-localized modes, 
regarded as zone-boundary breather spin waves, are pointed out by simulating the nonlinear difference 
equations describing the spin equilibrium positions in the chain, and are shown to exist irrespective 
of values of the model parameter. Domain-wall textures of the model are also numerically examined in terms of kink 
solitons and with regard to the parametrization. 
\end{abstract}


\maketitle

\section{\label{sec:level1}Introduction}
One-dimensional($1D$) magnets are simplest in their description while displaying at relatively low magnetic fields, a continuous symmetry 
related to spin precessions with respect to the chain axis~\cite{steiner1}. These systems have been widely investigated 
over the past years to both viewpoints of large and small-amplitude spin displacements and in connection with several real 
magnetic compounds, two most considered been $TMMC$ and $CsNiF_3$~\cite{steiner2,mikeska1,kjem,peter,wysin}. However, 
one-dimensional magnets are rather aboundant in 
the nature therefore implying quite various and often complex magnetic features e.g. structures. The first, $CsNiF_3$, 
is probably the 
best studied and relatively clear understanding of its structural properties has now been gained. In this 
compound, 
the Nickel ions align into a linear chain and are coupled through their magnetic moments giving rise to a $1D$ ferromagnetic 
system. In connection with this particular arrangement, magnetic moments experience an anisotropy field so called single-ion 
anisotropy and due to the crystalline structure. The crystalline anisotropy favors in-plane spin precessions where the rotation 
plane can be perpendicular to the chain axis. \\
When no magnetic field acts on the system, spins can freely rotate and their precessions lead 
either to large-amplitude or to small-amplitude exictations. Large precessions imply complete 
rotation hence $2\pi$ excitations which in theory are described as solitons. Solitons relate 
to domain walls or the elementary excitations of the collective spins. To a physical viewpoint, 
single domain walls represent local magnetic orders which, under thermal fluctuations, can nucleate 
giving rise to either short or long-range ordered ferromagnetic phase depending on the strength of the thermal fluctuations. 
As opposed to domain walls, 
small-amplitude magnetic excitations are more extended and dispersed hence do not necessarily cause complete breaking of 
a prevailing magnetic order. At the most 
they will 
promote order-disorder instabilities governed by spin-wave dispersions. However, at low enough 
temperatures, spin waves can coherently order into anharmonic or breather-like excitations. In fact, these breather-like structures 
form an hierarchy of varied localized modes in the ferromagnetic groundstate~\cite{mills} to which belong the intrinsic-self-localized(ISL) 
modes recently discussed~\cite{wallis,lai,takeno1} for both ferromagnetic and antiferromagnetic compounds. Since they 
describe excited states of the ferromagnetic groundstate, anharmonic localized modes can be looked on as classical 
anharmonic fluctuations in this groundstate. In general, such anharmonic fluctuations will exist irrespective of the 
wavector but depending mainly on the magnitude of the spin displacement. However, ISL modes are particular in that they 
need the wavector be exactly at the edge of the first Brillouin zone such that neighbour spins 
coherently oscillate anti-phase them and others but at the same frequency. In fact, all these coherent anharmonic objects 
extend our eyesight of the spin-wave phase of $1D$ magnets beyond the usual one restricting to linear phenomena. \\
In the presence of a uniform field perpendicular to the chain axis, the spin rotations become strictly planar and the 
ferromagnetic order almost confined along the chain axis. As a consequence the spin equilibrium and dynamical properties 
will strongly dependent on the magnitude of the applied field. While solitons in this last context still correspond to 
the progressive $2\pi$ rotation of spins about the easy axis, the region spanned by the spins is now 
determined by the field. Refering to the domain-wall texture, this region corresponds exactly to the spatial extension of a 
single domain hence the soliton characteristic width. Analytically, one finds that this region varies as the inverse 
square root of the magnitude of the applied field. Moreover, for the discrete planar model largely discussed 
in the literature, the magnetic field determines conditions for the propagation of soliton along the easy axis via a defined 
dispersion law as Takeno et al. pointed out~\cite{homma1,homma2}. \\
 Very recently, a $1D$ parametrized spin model appeared in the literature~\cite{kofane,nguenang1} and was shown to reduce to 
a parametrized sine-Gordon model previously suggested by Remoissenet and Peyrard~\cite{peyrard1,peyrard} and studied by 
several authors\cite{braun1,braun2,braun3,yemele,kevre}. In connection with this well-known model, the nonlinear dynamics of the 
$1D$ parametrized spin chain has been 
investigated~\cite{nguenang1,nguenang2} and reveal rich topological soliton features. In spite of a lacke of sound 
experimental contexts conforting such kind of parametrization, the existence of exact soliton solutions is a 
relevant source of motivation for an interest to this new model. Keep in mind that none of the current models 
i.e., neither the isotropic nor the planar ising or Heisenberg models has yet demonstrated ability to give complete 
account of experiments on the available magnetic systems, and will certainly not be more helpful for an 
understanding of the physics behind structures shown by the various magnetic alloys. In its original form, the 
parametrized spin model describes a $1D$ ferromagnetic chain subjected to an easy-plane single-ion anisitropy and a Zeeman 
energy with nonconvex dependence on the spin variable. While the 
author~\cite{kofane} relates this nonconvex dependence to some deformability, a more efficient and effective 
interpretation can be given in terms of a non-conventional crystalline structure such that the magnetic field cannot 
couple exactly to none of the spin components. In this context, the parametrization can stand as a relevant theoretical 
mean to include deviations from conventional crystalline structures. We therefore suspect such an approach to 
provide an interesting way to model some real complex structural magnets obtained e.g. by isotopic substitutions, 
magnetic dopings, alloyings, etc. While agreing with the original modelling, we retract from the explicit inclusion of 
the cristalline anisotropy energy. Indeed, the nonconvex 
Zeeman term is to our view already self-sufficient i.e. involves energies from the two main anisotropies.    \\
 We will focus on one single point i.e. the static properties of spins in this model, the ultimate goal being to check 
the feasability of such a parametrized spin model. In particular we will probe possible analytical weakness, namely whether 
some restrictions exist that prevent harmonic as well as anharmonic spin-wave and kink soliton phases. 
\section{\label{sec:level2}The spin-wave phase.}
\begin{figure}
\includegraphics[height=5cm]{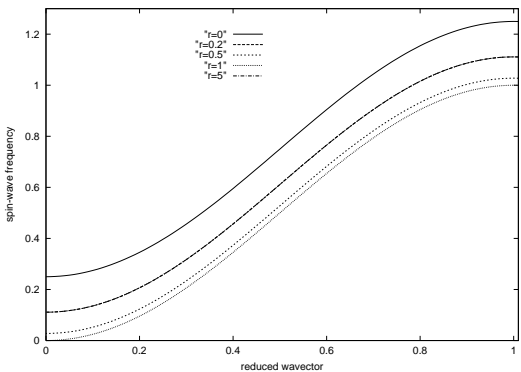}
\caption{\label{fig:gapa} The spin-wave dispersion spectrum, plotted in the reduced wavector $k \, a/\pi$.} 
\end{figure}

\begin{figure}
\includegraphics[height=5cm]{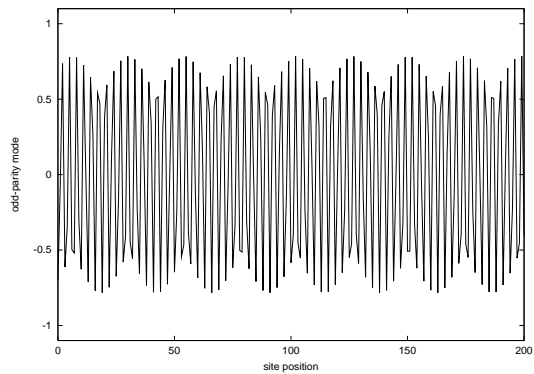}
\caption{\label{fig:solit1a} Odd-parity zone-boundary breather: ideal planar model $r=0$.}
\end{figure}

\begin{figure}
\includegraphics[height=5cm]{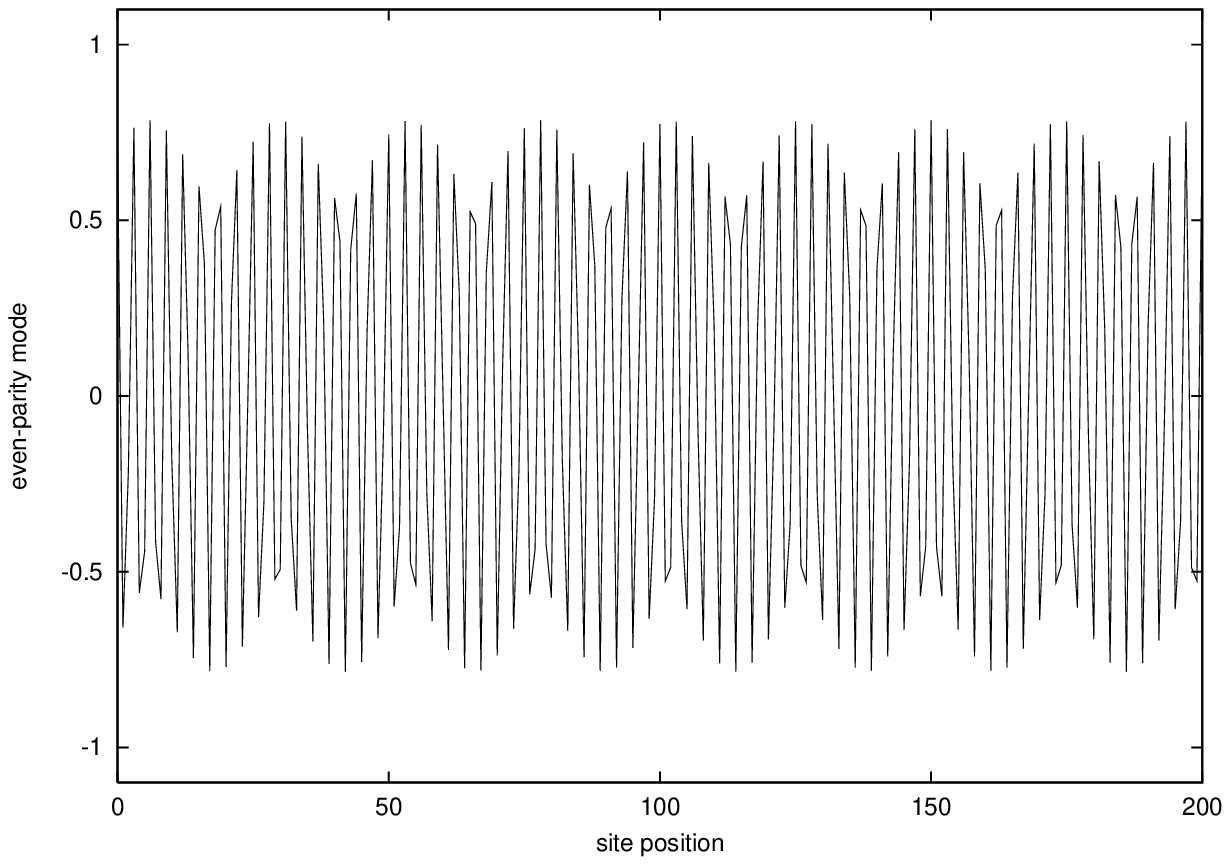}
\caption{\label{fig:solit1b} Even-parity zone-boundary breather: idela planar model $r=0$.}
\end{figure}

\begin{figure}
\includegraphics[height=5cm]{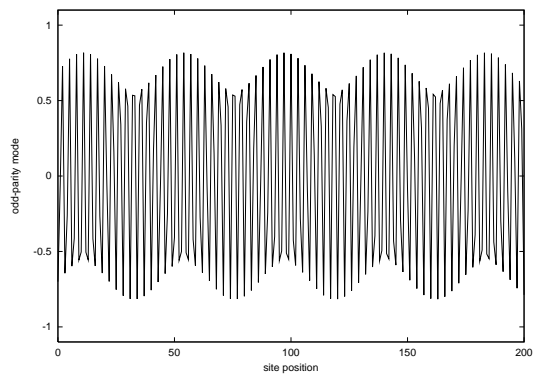}
\caption{\label{fig:solit2a} Odd-parity zone-boundary breather: $r=0.05$.}
\end{figure}

\begin{figure}
\includegraphics[height=5cm]{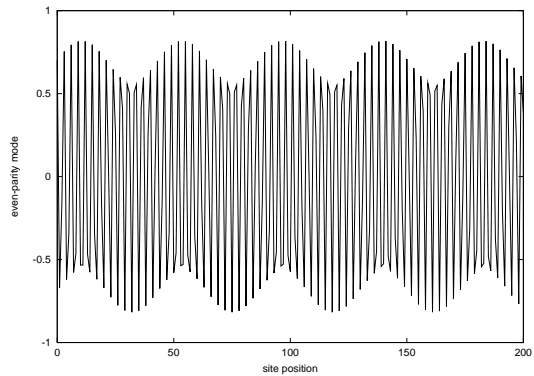}
\caption{\label{fig:solit2b} Even-parity zone-boundary breather: $r=0.05$.}
\end{figure}

Consider a $1D$ magnetic system with ferromagnetic interactions among the spins. We assume that the magnetic field does not couples exactly to 
none of the spin components such that for a field directed perpendicular to the z axis, the system Hamiltonian writes:
\begin{equation}
H= -2J\sum_n {{\bf s}_n {\bf s}_{n+1}} + h(r) \sum_n {\frac{1 - s_n^z}{1 + r^2 + 2r s_n^z}}, \label{hamil}
\end{equation}
where $J$ is the exchange coupling and $r$ some real parameter. $h(r)$ is a function defined as:
\begin{equation}
h(r)= h_o(1 - r)^2. \label{champ}
\end{equation}
where $h_o$ is the effective uniform field that, for a well-defined anisotropic crystalline structure with easy axis $z$, should 
act on the component $s_n^z$ of the spin vector ${\bf s}_n$. Compared with the original model~\cite{kofane}, the new form~(\ref{hamil}) retains only the nearest-neighbour exchange 
energy and the nonconvex anisotropy energy. The reason for such simplification follows from the remark that 
one can expand the nonconvex energy for weak values of $r$(or more 
precisely, of the ratio $2r/(1+r^2)$) and  up to the term in $(s_n^z)^2$, both Zeeman-like and 
single-ion-like contributions are generated. So in its 
present form, the parametrized model is self-consistent at least for the two main encountered anisotropies in $1D$ 
ferromagnets. Moreover, further expansions give new energy terms each either contributing of a new symmetry or enforcing an 
already existing symmetry. To this viewpoint, such an expansion 
gives rise to a generalized free-energy-like Hamiltonian in which the first term(i.e. the exchange energy) can be looked on 
as the distorsion energy. \\
Because of the product of the parameter $h_o$ with all terms of this expansion, a relevant 
question arises as to how symmetries get broken in the system. The answer resides in the fact that coefficients of this 
expansion are not of the same power in $r$. Therefore one can always define local fields characterized 
by different powers of the deformability parameter $r$, which then appears to rule symmetry-breaking phenomena in the 
non-conventional crystalline structure. In view of all these remarks, we will avoid calling the last term in~(\ref{hamil}) 
a Zeeman energy as in the previous works. Instead, we will refer to as a nonconvex anisotropy energy. \\
 As we are interested with the spin-wave phase of this model, we must distinguish two classes of spin waves characterized by 
two distinct amplitude regimes. However, both spin-wave regimes lie in the ferromagnetic groundstate, of which they describe 
classical fluctuations. One of these two regimes is dominated by small-amplitude deviations from the uniformly ordered 
states $s_n=S$, i.e. involves harmonic waves. This regime is best described using coherent state variables as given by the 
Holstein-Primakov transformation~\cite{holstein}. Expressing the spin variables $s_n$ as functions of these coherent-state 
variables, 
next expanding the nonconvex term in $r$ and grouping all quadratic terms in the coherent states we derive the 
following dispersion spectrum for the harmonic excitations:
\begin{equation}
\omega(q)= \omega_o sin^2(k\, a/2) + \frac{h_o}{\hbar}\left( \frac{1 - r}{1 + r} \right)^2, \hspace{.2in} \omega_o= \frac{8J}{\hbar},  \label{disp}
\end{equation}
Where $a$ is the lattice spacing and k the wavector. This dispersion relation reveals a gap in the groundstate of the spin-wave phase i.e. at $k=0$, of amplitude:
\begin{equation}
\Delta(r)= \frac{h_o}{\hbar}\left( \frac{1 - r}{1 + r} \right)^2, \label{gap}
\end{equation}
and which closes continuously as $r$ increases. According to figure~\ref{fig:gapa}, the field-induced gap causes uniform shift of the whole excitation spectrum such that not only the $k=0$ spin states of the ferromagnetic 
groundstate will be affected by its closing. 
Still, excitations involving the spin states near $k=0$ will be more prone to harmonic phenomena as opposed to excitations of the 
spin states lying close to the edge of the first Briulloiun zone, from which relatively large-amplitude phenomena are 
expected. Since these last excitations suggest possible anharmonic spin waves the classical spin motion should obey discrete nonlinear 
equations. Owing to the complex form of the model Hamiltonian, the technique of derivation of these equations deserve 
attention. We will follow a previous approach~\cite{lai} transforming the Hamiltonian as a function of scale(rising and lowering) variables 
$s_n^{\pm}= s_n^x \pm i s_n^y$, so that~(\ref{hamil}) can be rewritten: 
\begin{equation}
H = -J\sum_n{ \left[ s_n^+ s_{n+1}^- + s_n^- s_{n+1}^+ + 2 s_n^z s_{n+1}^z \right]} + \sum_n V(s_n^z, r), \label{mahlit11}
\end{equation}
\begin{equation}
V(s_n^z, r)=  h(r) \frac{1 - s_n^z}{1 + r^2 + 2r s_n^z}.     \label{hamil1}
\end{equation}
The next step is to apply the following canonical equation:
\begin{equation}
i\hbar \frac{d s_n^{\pm}}{d t}= \left[s_n^{\pm}, H \right], \hspace{.1in} \label{canon}
\end{equation}
which leads to the formal equation:
\begin{eqnarray}
i\hbar \frac{d s_n^{\pm}}{d t}&=& -2J \left[s_n^z\left(s_{n+1}^+ + S_{n-1}^+ \right) - s_n^+ \left(s_{n+1}^z + s_{n-1}^z \right) \right] \nonumber \\
 &+& \left[s_n^{\pm}, V(s_n^z,r) \right]. \label{canon1}
\end{eqnarray}
Let us expand the function $V(s_n^z, r)$ as a polynomial in $s_n^z$:
\begin{eqnarray}
V(s_n^z, r) &=& h(r)(1 - s_n^z) V_1(s_n^z, r), \nonumber \\
V_1(s_n^z, r)&=& \frac{1}{1 + r^2 + 2r s_n^z} \nonumber \\
             &=& \frac{1}{1+r^2} \sum_{p=1}{(-\lambda)^{p-1} (s_n^z)^{p-1}}, \nonumber \\
  \lambda&=& \frac{2r}{1+r^2}. \label{develop}
\end{eqnarray}
Proceeding term after term, we can establish that the commutator of the rising variable $s_n^+$ with the polynomial 
function $V_1$ of $s_n^z$ ultimately reduces to:
\begin{equation}
\left[s_n^+, V_1(s_n^z, r) \right]= \left[s_n^+, s_n^z\right] \frac{\partial V_1}{\partial s_n^z}. \label{partiel}
\end{equation}
With help of these transformations, the discrete equation~(\ref{canon1}) becomes more explicit. However, to derive this 
explicit form $s_n^+$ must first be expressed as a time-dependent quantity. In this goal we define:
\begin{equation}
s_n^+ = s_n(t) e^{i\left(kna - \Omega t\right)}. \label{fourier}
\end{equation}
Inserting~(\ref{fourier}) in~(\ref{canon1}) and~(\ref{partiel}),  and next separating real and imaginary parts of the 
resulting equation, we arrive at two new equations i.e.:
\begin{eqnarray}
\frac{d s_n(t)}{d \tau} &=& - \sqrt{1 - \vert s_n(t) \vert^2} \left[ s_{n+1}(t) - s_{n-1}(t) \right] \sin(k), \nonumber \\
 \tau &=& \frac{2J}{\hbar}t, \label{dynam}
\end{eqnarray}
\begin{eqnarray}
\bar{\Omega} s_n &=& s_n \left[\sqrt{1 - \vert s_{n+1} \vert^2} + \sqrt{1 - \vert s_{n-1} \vert^2} \right] \nonumber \\
           &-& \sqrt{1 - \vert s_n \vert^2}\left[s_{n+1} + s_{n-1}\right] cos(k) \nonumber \\
	   &+& \frac{h(r)}{2J} \frac{s_n}{\left[1 + r^2 + 2r \sqrt{1 - \vert s_n \vert^2} \right]^2}, \nonumber \\
\bar{\Omega}&=& \frac{\hbar}{2J}\Omega, \hspace{.2in} \vert s_n \vert^2 + (s_n^z)^2= S^2,   \label{eqat1}
\end{eqnarray}  
where we assume $S=1$. Equations~(\ref{dynam}) and~(\ref{eqat1}) form the complete set governing the dynamics of anharmonic 
spin-wave excitations of the parametrized model. However, to keep our primary goal we will concentrate on the static 
part of this set. Recall that this static part is most relevant for the equilibrium properties of the ferromagnetic 
groundstate in the presence of the anharmonic classical(spatial) fluctuations. In this context, the discrete static 
equations~(\ref{eqat1}) provide reccurence relations between positions of neighbour spins fluctuating 
about their equilibra. The presence of $\Omega$ in these equations, which is the frequency at which spins 
fluctuate oscillating in time, traduces an account of adiabatic spin fluctuations thus meaning quite rich 
equilibrium properties. Notably, ISL modes will occur as anharmonic classical fluctuations 
corresponding to the spin-wave phase in which $\Omega$ gets closer and closer to the cut-off frequency 
$\omega(k=\pi/a)$, such that spins of same partity(e.g. even-site spins) coherently oscillate in phase being completely 
decoupled from the spins of opposite parity(odd-site spins). So to say we expect two co-existing anharmonic modes at this specific 
wavector, leading to two degenerate localized spin-wave excitations in the ferromagnetic groundstate. 
Here the degeneracy relates to the fact that both localized spin-wave modes fluctuate at a common frequency though 
being completely decoupled one from the other. Instructively, the phenomenon just described looks quite like the nonlinear 
self-localized objects predicted by Chen 
and Mills~\cite{chen} in photonic band-gap crystals, and extended to other lattice systems e.g. atomic 
systems~\cite{takeno2,kivshar1,kiselev} 
and nonlinear transmission lines~\cite{essimbi1,essimbi2}. The most attracting feature in these particular anharmonic modes is their 
common breather-like shape profiles. Stress that by breather we understand an harmonic wave 
hidden in a spatial nonlinear topological shell. Being robust by virtue of its topological feature, we readily expect this spatial 
modulation to increase lifetime of the harmonic wave, namely by fixing the frequency of its "breath" inside the 
shell. In this respect, the most common belief supported by numerous theoretical results is that the breather frequency will depend on 
the width of the spatial topological shell, which often is one of the two most popular topological solitons i.e. kink or pulse. \\
Let us now put the previous discussions on a firm ground by carrying out numerical simulations of the discrete nonlinear 
equation~(\ref{eqat1}). As a starting note, following these previous discussions we can also term ISL modes zone-boundary 
breathers. Indeed such a description reflects rather well both their well-known waveshapes, and the fact that they involve 
wavectors lying close to the boundary of the first Brillouin zone($k=\pi/a$). In connection with this last requirement, it 
is useful to clearly define the interval of allowed frequencies as given by the dispersion relation~(\ref{disp}). From the 
condition $0 \leq k \leq \pi/a$ the dispersion relation yields: 
\begin{equation}
h \left( \frac{1-r}{1+r} \right)^2 \leq \Omega \leq \omega_o + h \left( \frac{1-r}{1+r} \right)^2, \label{contraint}
\end{equation}   
As it turns out, placing ourselves near the edge of the first Brillouin zone means choosing $\Omega$ close to the upper 
limit of the constraint~(\ref{contraint}). Thus, once we fix the ratio $h_o/J$ an appropriate choice of the zone-boundary 
breather frequency follows. In our simulations, $h_o/J$ is kept fixed at 0.1 which allows us selecting a frequency of about 4($J \equiv \hbar \equiv 1$) or slightly less. Last, periodic boundary conditions will be assumed and applied such that neighbouring spins start 
fluctuating simultaneously with opposite amplitudes. Numerically, this amounts to set $s_0^{ep}= - s_0^{op}$ where 
superscripts "ep" and "op" refer to even-parity and odd-parity modes, respectively.  \\
On figures~\ref{fig:solit1a} and~\ref{fig:solit1b}, we illustrate shapes of the odd and even parity modes for the conventional 
planar model $r=0$. We clearly see that their maxima are uniformly shifted which is a signature of their anti-phase oscillations. 
Otherwise, the twin character of their shapes constitutes a sound proof of a ferromagnetic ordering in the groundstate of the 
conventional planar model. 
An instructive step following this previous relevant one is to assume weak values of r and check the becoming of the zone-boundary 
breathers in figure~\ref{fig:solit1a} and~\ref{fig:solit1b}. Figures~\ref{fig:solit2a} and~\ref{fig:solit2b} show numerical 
results for $r=0.05$. 
As these figures suggest, a nonzero value of $r$ does not changes the physics contained in the conventional planar model. 
Namely, the groundstate remains ferromagnetically ordered and manifestly as stable as the groundstate of the conventional 
planar model. In fact, $r$ 
only affects shape profiles of the odd and even parity zone-boundary breathers but not their existence. More explicitely, 
an increase of r causes their spatial extensions(widths) to increase while preserving their twin structures. From this step we can already conclude that the parametrized model allow 
smooth variation of characteristic magnetic features from the ideal limit corresponding to the usual planar model. However, to 
permit global quantitative and qualitative appreciations we will display numerical solutions  for larger values of r. Thus, figures~\ref{fig:solit3a},~\ref{fig:solit3b},~{\ref{fig:solit3c},~\ref{fig:solit3d} 
and~\ref{fig:solit3e} are some representative shape profiles. As we are aware of the twin character of the two modes, only odd-parity solutions are shown 
in this last set of figures.
 \begin{figure}
\includegraphics[height=5cm]{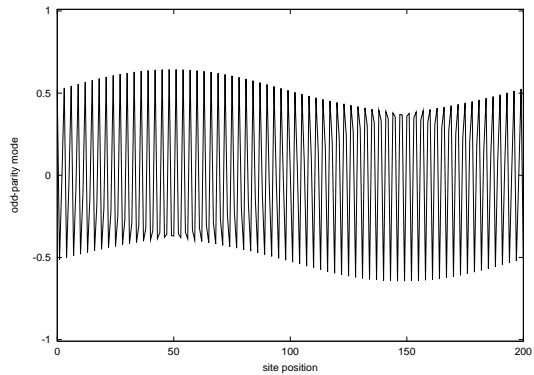}
\caption{\label{fig:solit3a} Shape of the zone-boundary breather: $r=0.2$.}
\end{figure}
\begin{figure}
\includegraphics[height=5cm]{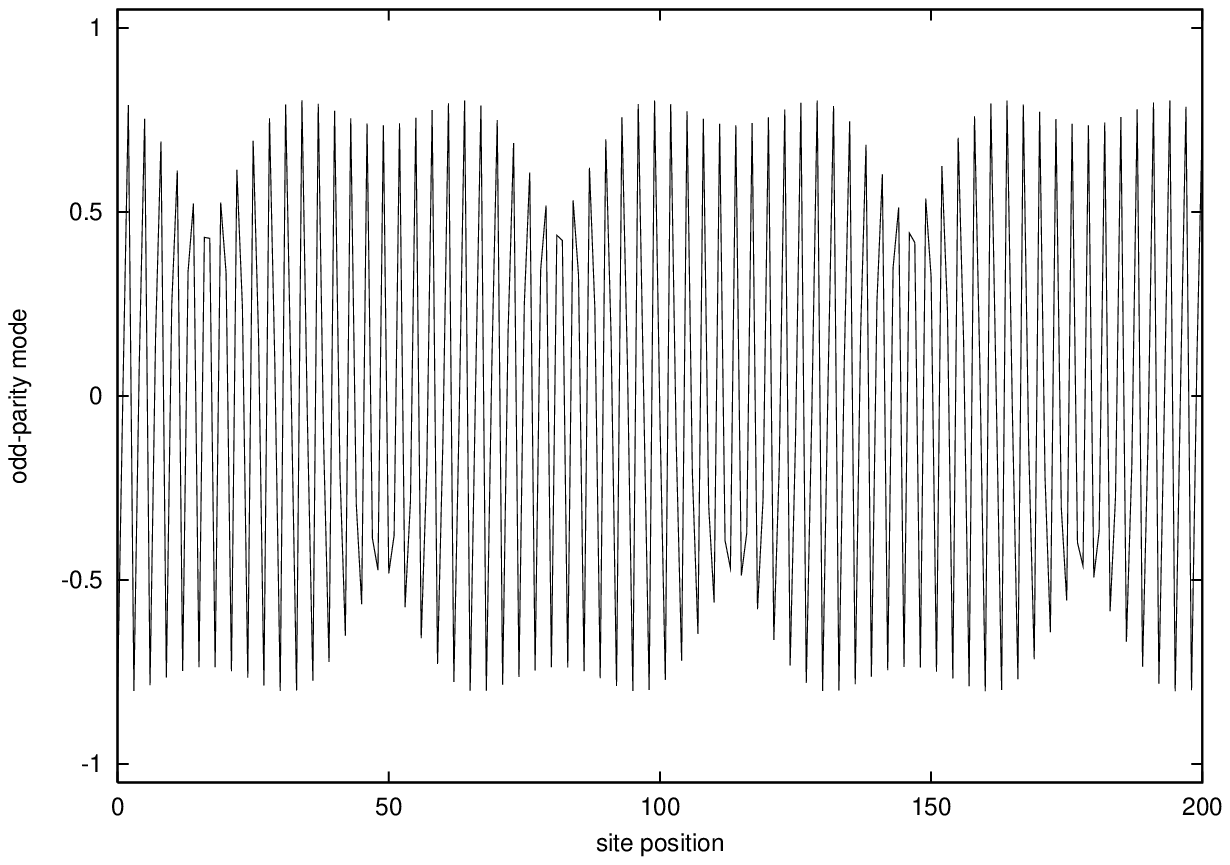}
\caption{\label{fig:solit3b} Shape of the zone-boundary breather: $r=0.5$.}
\end{figure}
\begin{figure}
\includegraphics[height=5cm]{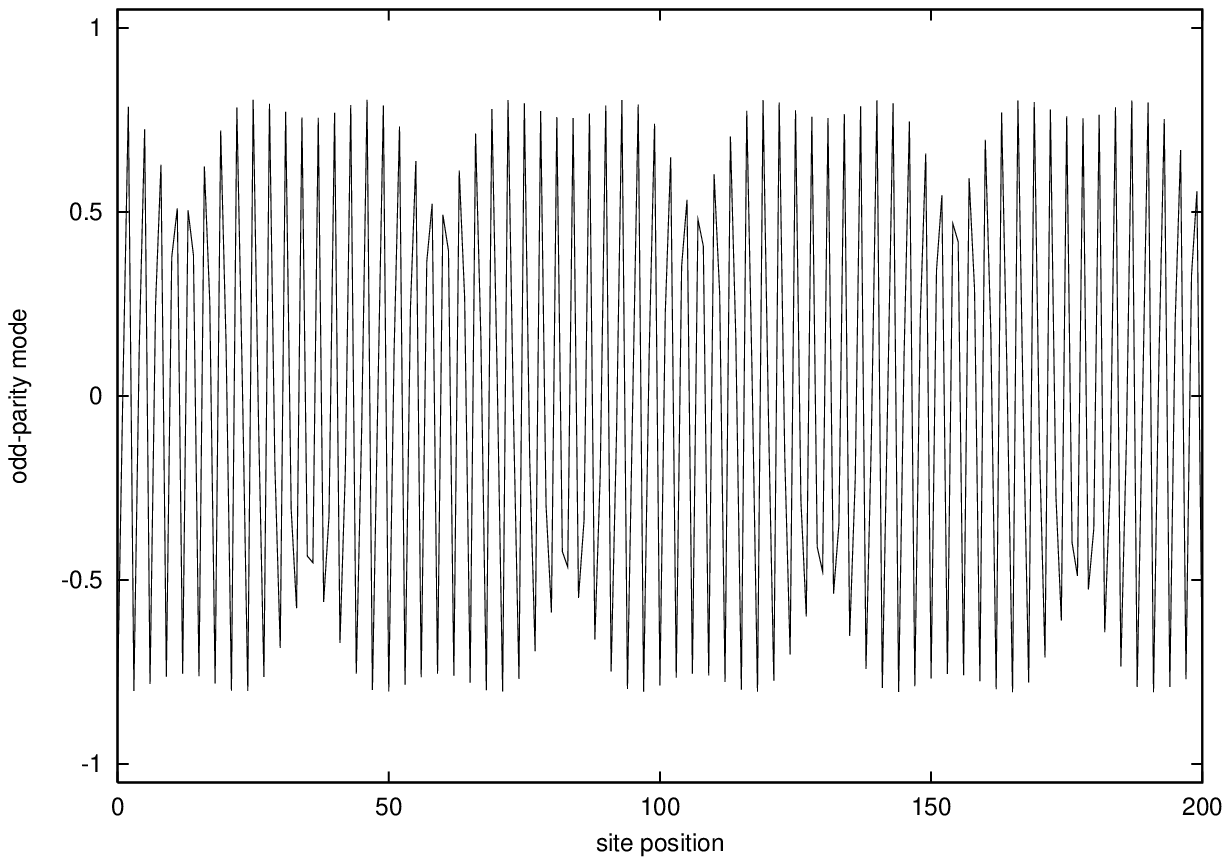}
\caption{\label{fig:aolit3c} Shape of the zone-boundary breather: $r=0.8$.}
\end{figure}
\begin{figure}
\includegraphics[height=5cm]{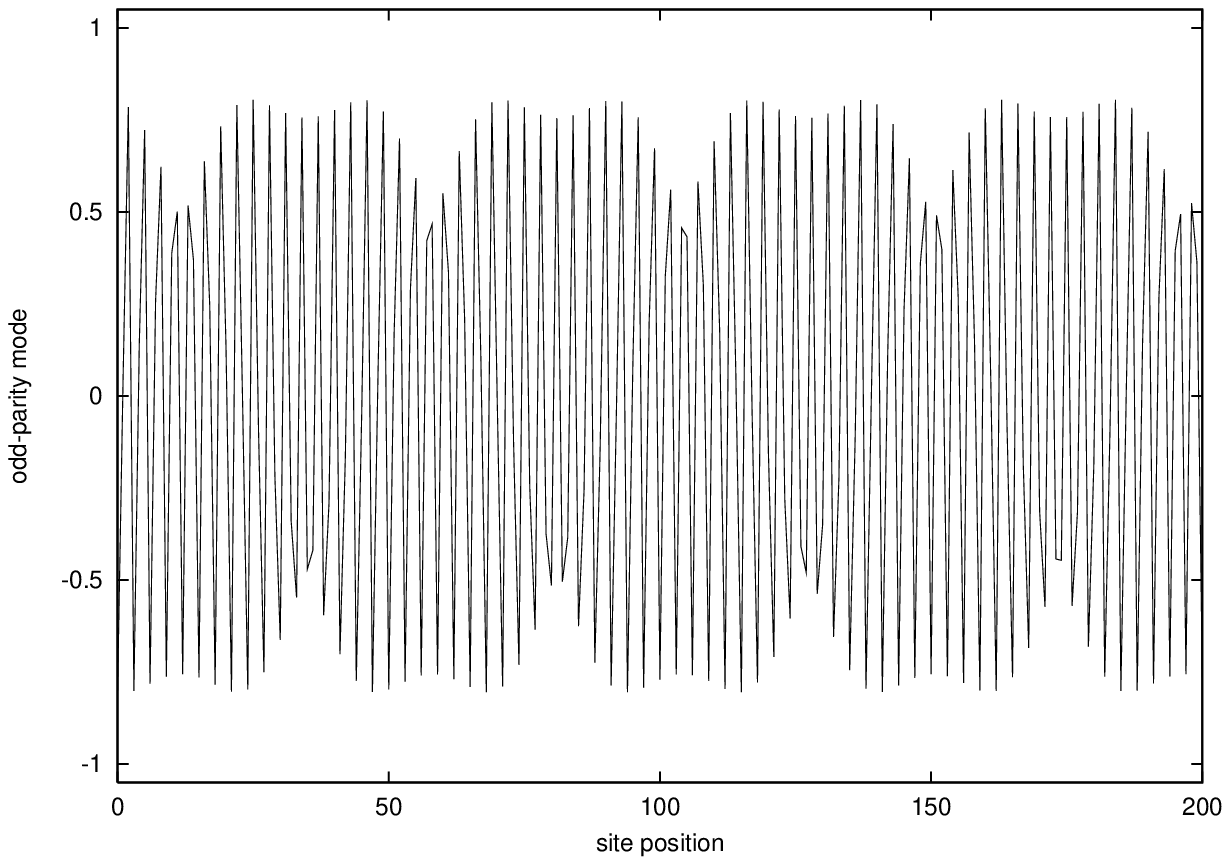}
\caption{\label{fig:solit3d} Shape of the zone-boundary breather: $r=1$.}
\end{figure}
\begin{figure}
\includegraphics[height=5cm]{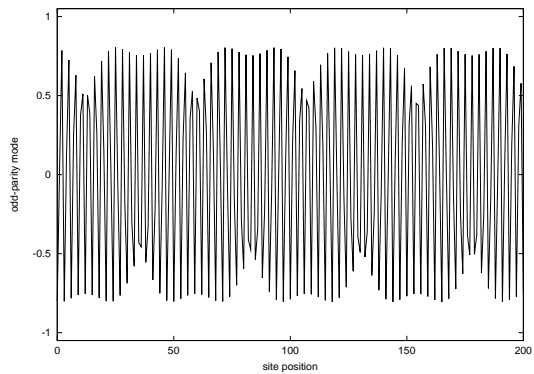}
\caption{\label{fig:solit3e} Shape of the zone-boundary breather: $r=10$.}
\end{figure}
\section{\label{sec:level3}Textures of domain walls.}
Soliton solutions of the Remoissenet-Peyrard~\cite{peyrard1,peyrard} model, which we recall is currently the most celebrated generalized 
sine-Gordon model, have been thoroughly investigated~\cite{braun1,braun2,braun3,yemele,kevre} both analytically and numerically. Various applications have so far been envisaged 
extending from Biophysics to crystal growths through magnetic excitations in $1D$ magnets. The discrete version of the sine-Gordon 
model has been introduced two decades ago~\cite{homma1,homma2} as a model for DNA conformational transitions. This model was next extended to various other contexts but the most 
evident system in which it naturally arises is the $1D$ anisotropic Heisenberg model. Here we will see that the parametrized model~(\ref{hamil}) gives 
rise to the parametrized counterpart of this model, namely the discrete Remoissenet-Peyrard model. In this goal, we assume the usual parametrization of spin variables
 using Euler angles i.e.:
\begin{equation}
s_n^x= sin \, \theta_n \, cos \, \phi_n, \hspace{.1in} s_n^y= sin \, \theta_n \, sin \phi_n, \hspace{.1in} s_n^z= cos \, \phi_n, \label{angles}
\end{equation}
where $\theta_n$ and $\phi_n$ are more precisely the local variables describing rotations of the $n^{th}$ spin. When these two 
angular variables assume arbitrary values, spins undergo precessional out-of-plane motions i.e. rotations out of the plane perpendicular to the chain axis. 
Nevertheless, we can confine spin motions in a fixed plane perpendicular to the chain axis by setting $\theta_n=\pi/2$. With this constraint, the total Hamiltonian~(\ref{hamil}) 
gives rise to the following nonlinear difference equations.
\begin{eqnarray}
 sin(\phi_{n+1} - \phi_n) - sin(\phi_n - \phi_{n-1}) \nonumber \\
   - \frac{1}{\ell_o^2} \frac{sin \, \phi_n}{\left[ 1 + r^2 + 2rcos \, \phi_n \right]^2} = 0, \label{discret}
 \end{eqnarray}
 \begin{equation}
 \ell_o^2= \frac{2J}{h(1 - r^2)^2}.  \label{length}
\end{equation}
These equations describe the texture of the space spanned by the $n^{th}$ spin rotation in the presence of two 
nearest-neighbour spins. The energy density associate to this spatial texture is readily obtained as: 
\begin{eqnarray} 
 e_n&=& cos(\phi_{n+1} - \phi_n) + cos(\phi_n - \phi_{n-1}) \nonumber \\
    &-& \frac{h(r)}{2J} \frac{1 - cos \, \phi_n}{1 + r^2 + 2rcos \, \phi_n}, \label{energy}
\end{eqnarray}
The continuum version of~(\ref{discret}), i.e. the Remoissenet-Peyrard equation, admits exact topological kink solutions. We can therefore anticipate 
predicting exact soliton solutions for the discrete counterpart. In this viewpoint,~(\ref{energy}) is nothing else but the energy density needed 
for the kink soliton to keep its topological shape. \\ 
We have simulated the discrete equation~(\ref{discret}) as well as the energy density~(\ref{energy}). On 
figures~\ref{fig:solit4a},~\ref{fig:solit4b},~\ref{fig:solit4c},~\ref{fig:solit4d},~\ref{fig:solit4e},~\ref{fig:solit4f},~\ref{fig:solit4g} 
and~\ref{fig:solit4h}, we plot results for $r=0.0$, 0.2, 0.5, and 10, respectively. The most stricking feature in all the figures representing solutions 
is the well-formed kink shape. Moreover, the kink-lattice structure is well developped and apparently exists for a rather broad range of values of the model parameter. 
It is interesting to note that no kink solution exists for $r=1$, which is predictable as there the "nonlinear force field" vanishes. To our knowledge, this is the only singular 
point of the model. In this respect, the usual restriction confining r between -1 and 1 is in deep contradiction with our own numerical findings. According to our simulations, 
the most remarkable fact at large values of r is the negative energy density. However, the kink-lattice structure still 
shows apparent stability as one can note on figure~\ref{fig:solit4g}. Morever, both the dispersion relation~(\ref{disp}) 
and the anharmonic spin-wave modes of figure~\ref{fig:solit3e} showed that the ferromagnetic groundstate was not so drastically 
affected by large values 
of $r$. Instead, what we noted was some cycle of variation of physical parameters with increasing $r$, for instance a close 
look on figure~\ref{fig:gapa} reveals that the dispersion curve for 
$r=5$ has a gap whereas the one corresponding to $r \sim 1$ or close is gapless. Similar behaviour also shows up in the variation of the shape of the ISL modes. Namely, figure~\ref{fig:solit3e} suggests a 
waveshape not so different from those found for values of $r$ less than 1. The negative value of the energy density as given in figure~\ref{fig:solit4h} 
also relate to this particular variations of the model characteristic parameters at large $r$. However, for this last case, 
one will easily check 
that the negative energy density is compensated by increasing the ratio $h_o/J$. We can therefore expect similar corrections on the previously 
simulated quantities by judicious adjustments of this ratio. 
\begin{figure}
\includegraphics[height=5cm]{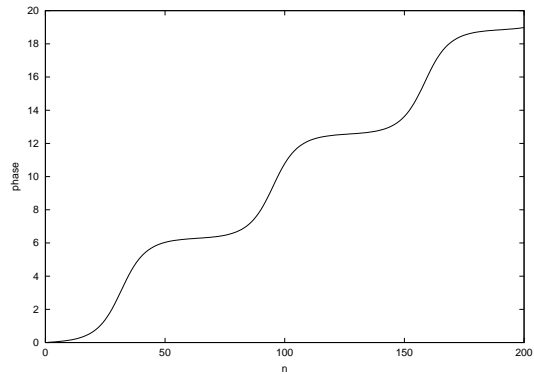}
\caption{\label{fig:solit4a} Kink lattice of the planar Heisenberg model($r=0.0$).}
\end{figure}
\begin{figure}
\includegraphics[height=5cm]{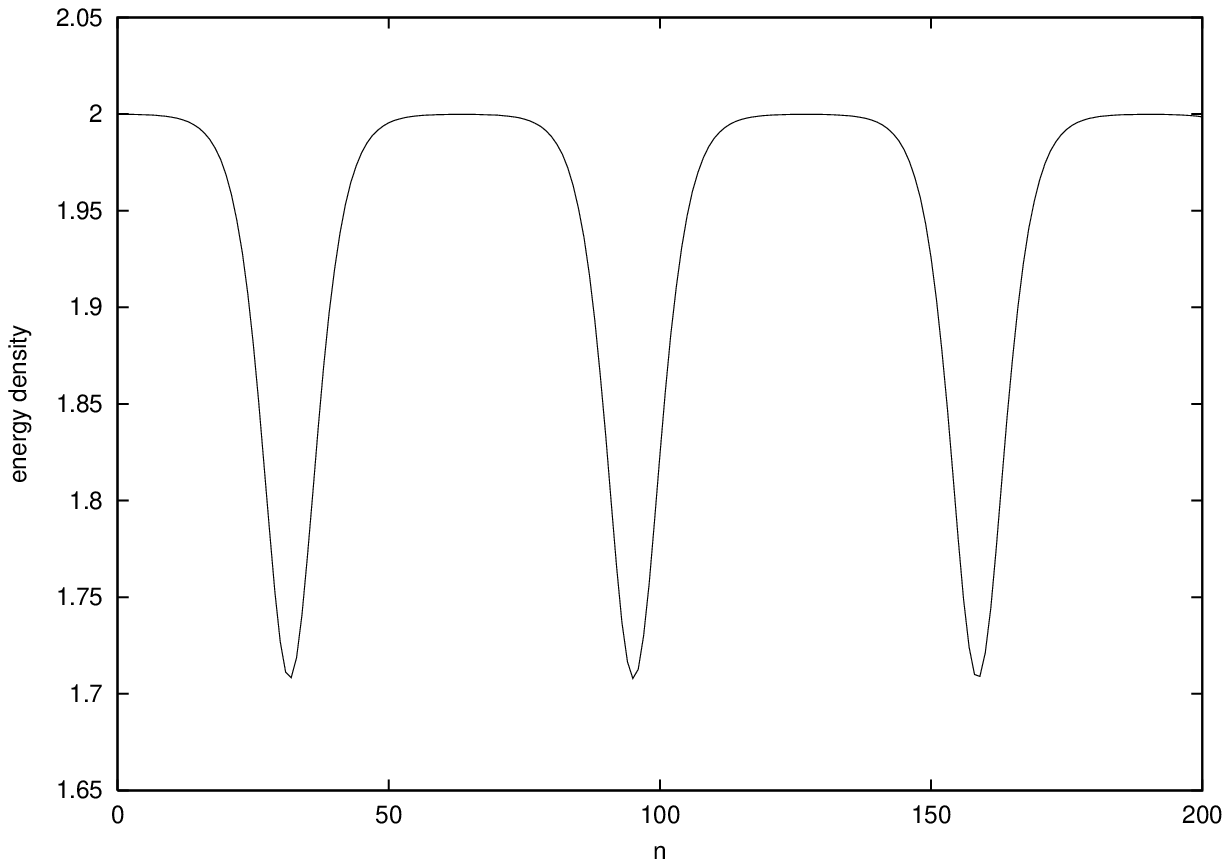}
\caption{\label{fig:solit4b} Kink energy density for $r=0$.}
\end{figure}
\begin{figure}
\includegraphics[height=5cm]{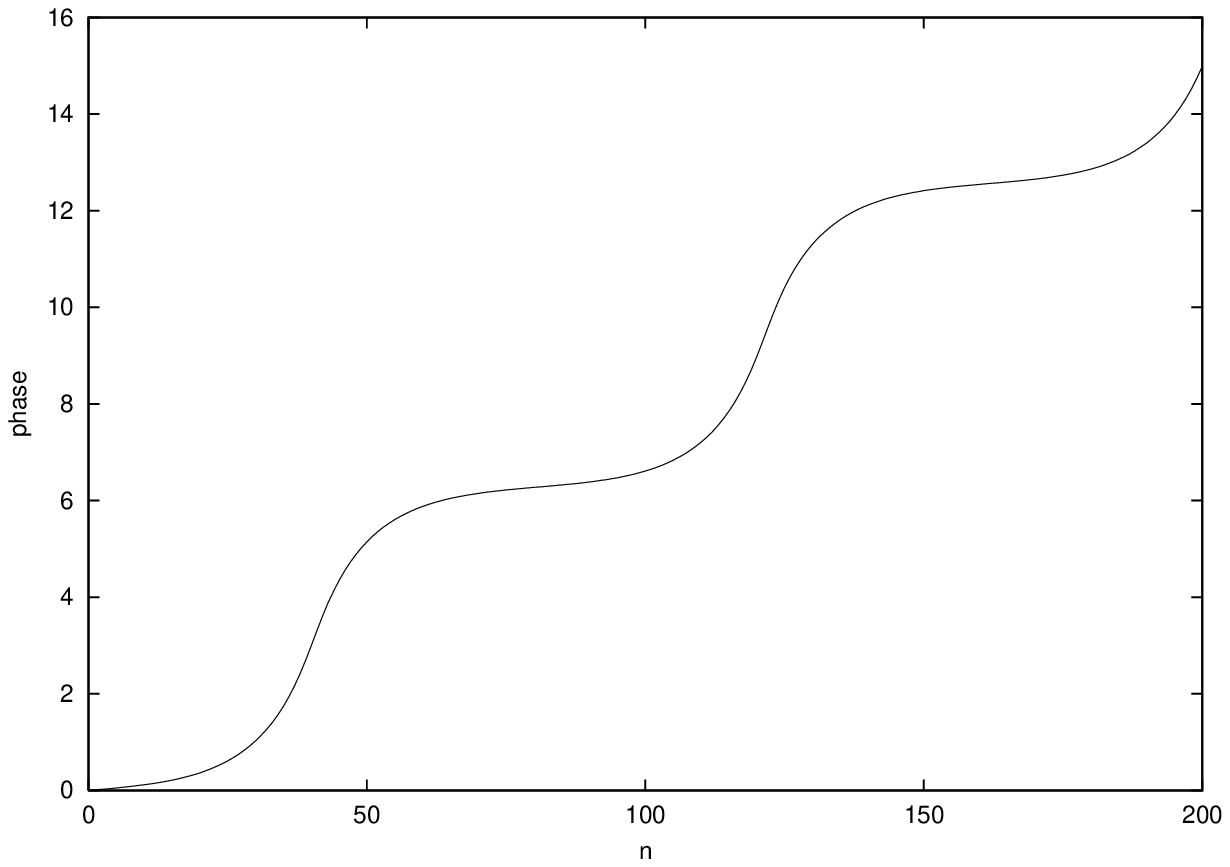}
\caption{\label{fig:solit4c} The kink lattice for $r=0.2$.}
\end{figure}
\begin{figure}
\includegraphics[height=5cm]{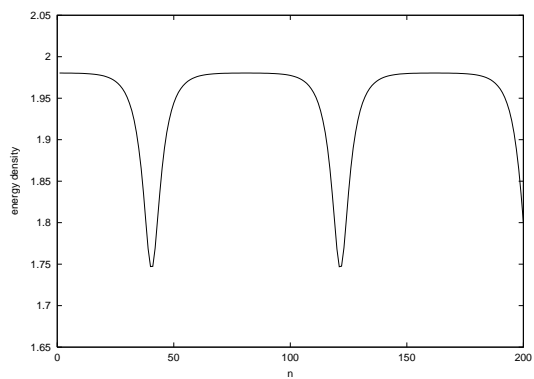}
\caption{\label{fig:solit4d} Kink energy density $r=0.2$.}
\end{figure}
\begin{figure}
\includegraphics[height=5cm]{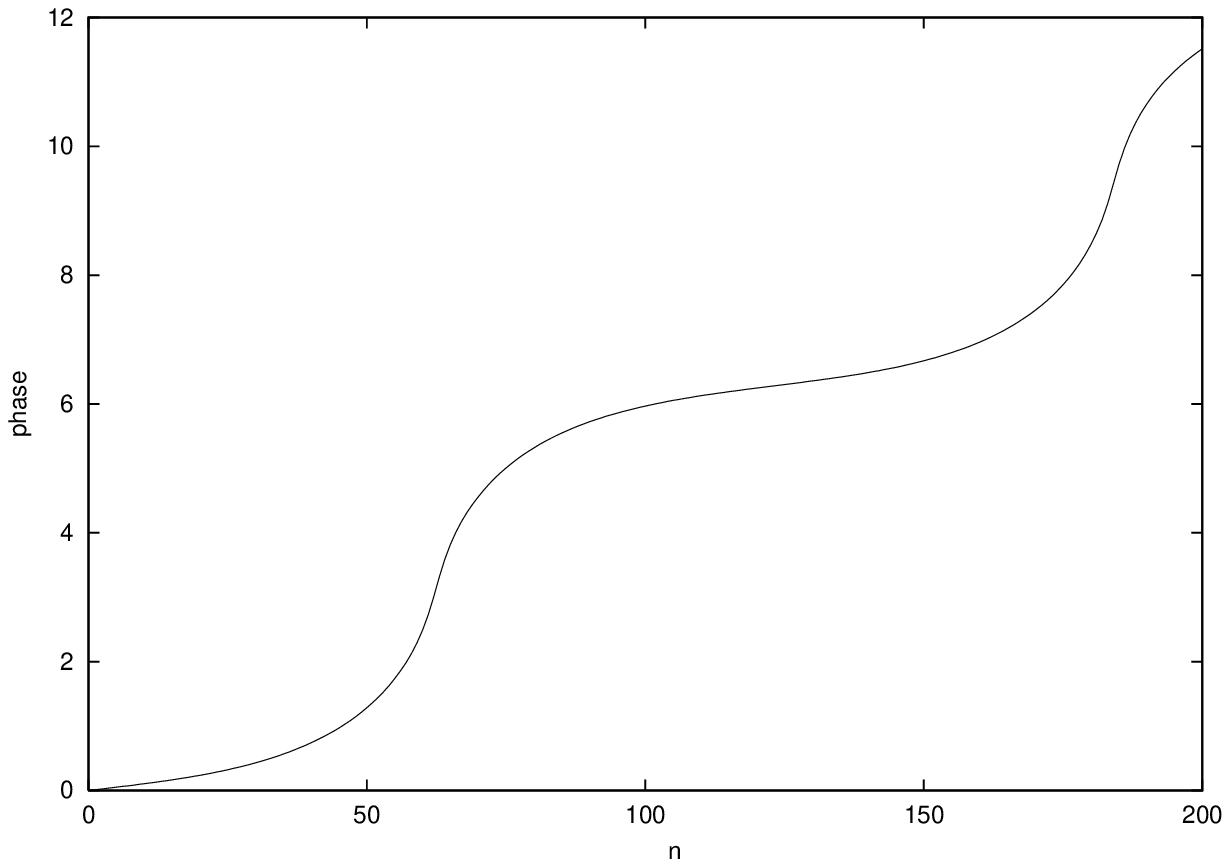}
\caption{\label{fig:solit4e} The kink lattice for $r=0.5$.}
\end{figure}
\begin{figure}
\includegraphics[height=5cm]{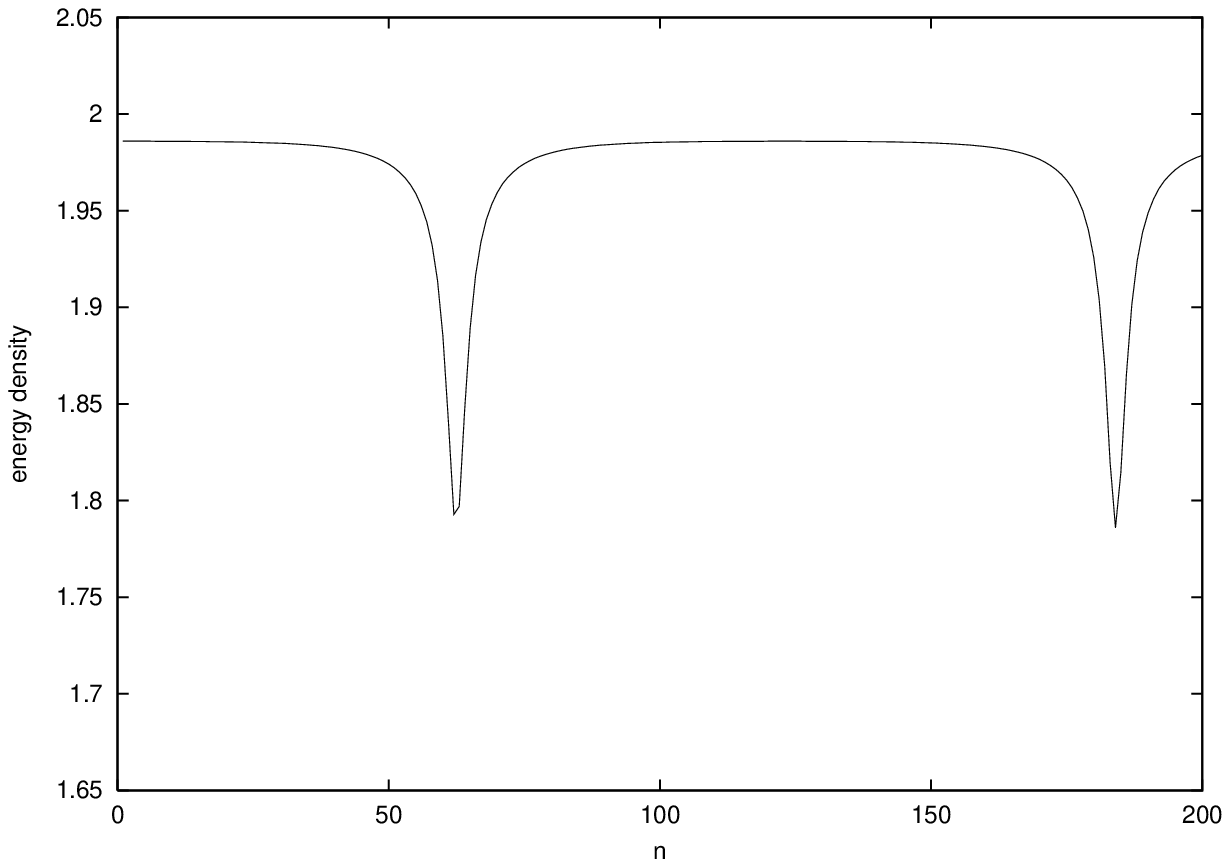}
\caption{\label{fig:solit4f} Kink energy density for $r=0.5$.}
\end{figure}
\begin{figure}
\includegraphics[height=5cm]{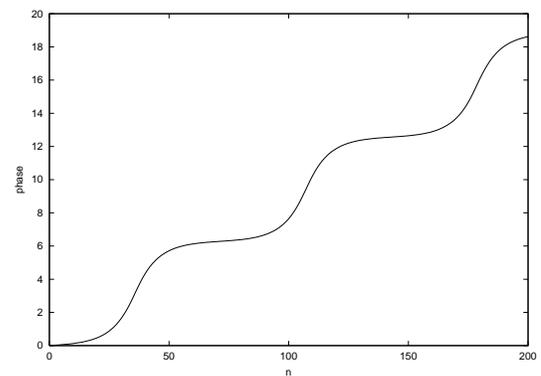}
\caption{\label{fig:solit4g} The kink lattice for $r=10$.}
\label{fig:solit4g}  
\end{figure}
\begin{figure}
\includegraphics[height=5cm]{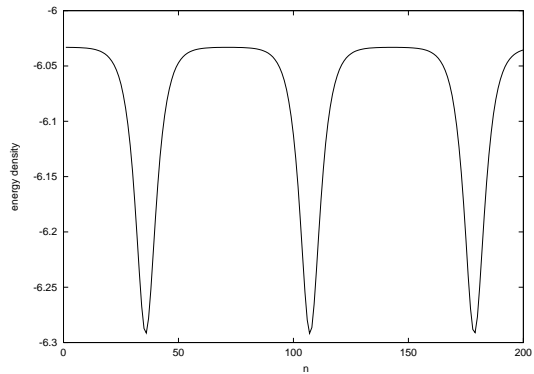}
\caption{\label{fig:solit4h} Kink energy density for $r=10$.}
\end{figure}
\section{\label{sec:level4} Concluding remarks.}
We have explored the static properties of a parametrized model as a probe of its ability to 
help describing the smooth variations of characteristic magnetic properties as it can occur in some ferromagnetic alloy series. 
In particular, we have in mind ferromagnetic systems with planar anisotropic structures but showing slight deviations from 
the conventional planar model, in which an applied magnetic field couples exactly to the spin component along the anisotropy axis. 
The parametrized model introduced as illustration displays rather rich features to both 
viewpoints of its spin-wave phase and domain-wall phase. Recall that this parametrized model is actually an improved version of a 
previously suggested spin model~\cite{kofane,nguenang1}. While the current version is more rigourous i.e. more appropriate for 
ferromagnetic systems often involving single crystalline anisotropy, the original version which is more general can become quite powerful for systems in which both an 
easy-axis and an easy-plane anisotropies co-exist. Still, as far as we know, simultaneous crystalline anisotropies of this kind 
are more probable in ferrimagnets or antiferromagnets but less in a true ferromagnetic spin system.

\end{document}